\documentclass[a4paper,12pt]{article}
\usepackage{mytitle}

\usepackage[T1]{fontenc}
\usepackage{fullpage}
\usepackage{hyperref}
\hypersetup{pageanchor=false}
\usepackage{breakurl}

\usepackage{axodraw2}

\usepackage{amsmath}

\usepackage{listings}
\usepackage{listings-form}
\usepackage{color}
\definecolor{DeclarationColor}{rgb}{0.0,0.5,0.1}
\definecolor{ExcutableColor}{rgb}{0.1,0.2,0.8}
\definecolor{ModuleColor}{rgb}{0.8,0.1,0.1}
\definecolor{DirectiveColor}{rgb}{0.8,0.1,0.8}
\definecolor{FunctionColor}{rgb}{0.0,0.5,0.9}
\definecolor{StringColor}{rgb}{0.8,0.4,0.0}
\definecolor{CommentColor}{rgb}{0.5,0.5,0.5}
\lstnewenvironment{verbatim}[1][]{\lstset{#1}}{}

\newcommand{\FORM}{\textsc{Form}}
\newcommand{\TFORM}{\textsc{TForm}}
\newcommand{\cmd}{\bgroup\catcode`\$=12\catcode`\#=12\catcode`\_=12\relax
                  \cmdwitharg}
\newcommand{\cmdwitharg}[1]{\textbf{\ttfamily #1}\egroup}

\begin{document}

\title{\FORM{} version 4.2}
\author[a,b]{B. Ruijl}
\author[a]{T. Ueda}
\author[a]{J.A.M. Vermaseren}
\affiliation[a]{%
  Nikhef Theory Group, \\
  Science Park 105, 1098 XG Amsterdam,
  The Netherlands}
\affiliation[b]{%
  Leiden Centre of Data Science, Leiden University, \\
  Niels Bohrweg 1, 2333 CA Leiden,
  The Netherlands}
\abstract{%
  We introduce \FORM{} 4.2, a new minor release of the symbolic manipulation toolkit. We demonstrate several new features,
  such as a new pattern matching option, new output optimization, and 
  automatic expansion of rational functions.
}

\maketitle

\section{Introduction}
\label{sec:introduction}

We introduce a new version of the symbolic manipulation toolkit \FORM{}~\cite{Vermaseren:2000nd,Kuipers:2012rf},
named \FORM{} 4.2. \FORM{} aims to be a high-performance symbol manipulator, even in cases where there are billions of
terms that take up several terabytes of disk space.

This release is an update from \FORM{} 4.1, which was released in 2013. In the latest release we have fixed 
more than 50 bugs and have introduced more than 20 new features. 
For an overview of all changes made from \FORM{} 4.1 to \FORM{} 4.2, please see the Changelog~\cite{Changelog}.
The driving force of the development of new features and improvements in \FORM{}
is the use of the program in actual research projects.
Many of the new features in \FORM{} 4.2 were inspired by
the use in \textsc{Forcer}~\cite{FORCER} but should be useful in other environments as well.
The correctness, efficiency and limitations of the new features have been
extensively tested through its physics applications.

The latest version of \FORM{} can be obtained from GitHub~\cite{Repository}, where installation instructions can also be
found.
The installation of \FORM{}, including the threaded version \TFORM{}~%
\cite{Tentyukov:2007mu}, on \texttt{x86-64} machines with normal
Linux distributions should be rather straightforward.

Whenever new features are introduced, users will resort to creative 
applications that are beyond the imagination of the designers. This can 
mean that certain restrictions, that were originally thought to never cause 
problems, can become an obstacle to such innovative use. Hence, if the user 
encounters issues, please report them to the GitHub Issue Tracker~\cite{Issues}.
One can also search for known issues in the Issue Tracker.

The layout of this work is as follows. In section \ref{sec:features} we show some new features and in section
\ref{sec:summary} we give a summary.

\section{Notable new features}
\label{sec:features}

In this section we present some notable new features of \FORM{} 4.2.
For more details of all features available in \FORM{} 4.2, please see the reference manual~\cite{Manual}.

\subsection{Generating all matches}

The \cmd{Identify} statement, often abbreviated as \cmd{id}, is one of core
statements of \FORM{}.
This statement tries to match a given pattern and
performs algebraic replacements.
\FORM{} version 4.2 introduces a new option \cmd{all}, which tries to generate
\emph{all} possible matches instead of just the first. For example,
\begin{verbatim}
CF v,f,s;
L F = v(1,2,3,4);
id all v(?a,?b) = f(?a)*s(?b);
Print +s;
.end
\end{verbatim}
gives the following result:
\begin{verbatim}
   F =
      + f*s(1,2,3,4)
      + f(1)*s(2,3,4)
      + f(1,2)*s(3,4)
      + f(1,2,3)*s(4)
      + f(1,2,3,4)*s
     ;
\end{verbatim}

The \cmd{id all} statement can be used in many different scenarios. One
is to find all automorphisms (symmetries) of a graph, described by the vertex structure
\cmd{vx}, which have (undirected) momenta assigned to all edges connected to the
vertex as function arguments~\cite{jvLL2016}.

\begin{figure}
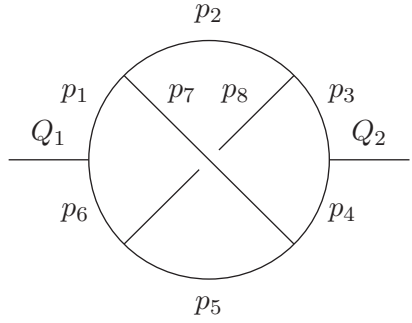

\centering
\begin{axopicture}{(90,90)(-40,-50)}

\CCirc(0,0){45}{Black}{White}
\Line(-75,0)(-45,0)
\Line(75,0)(45,0)
\Line(31.819,31.819)(-31.819,-31.819)
\CCirc(0,0){5}{White}{White}
\Line(-31.819,31.819)(31.819,-31.819)
\Text(-60,10) {\small $Q_1$}
\Text(60,10) {\small $Q_2$}
\Text(-50,25) {\small $p_1$}
\Text(50,25) {\small $p_3$}
\Text(0,55) {\small $p_2$}
\Text(0,-55) {\small $p_5$}
\Text(-10,25) {\small $p_7$}
\Text(10,25) {\small $p_8$}
\Text(-50,-20){\small $p_6$}
\Text(50,-20){\small $p_4$}
\end{axopicture}
\caption{Topology for which we find all automorphisms with \cmd{id all}.}
\label{fig:no}
\end{figure}

In figure~\ref{fig:no} we show a graph for which we find all eight
automorphisms using the following code:
\begin{verbatim}
V Q,Q1,Q2,p1,...,p8;
CF map,vx(s);  * note that the vertex structure vx is symmetric

L F = vx(Q1,p1,p6)*vx(p1,p2,p7)*vx(p2,p3,p8)*
      vx(p3,p4,Q2)*vx(p4,p5,p7)*vx(p5,p6,p8);

id all vx(Q1?,p1?,p6?)*vx(p1?,p2?,p7?)*vx(p2?,p3?,p8?)*
       vx(p3?,p4?,Q2?)*vx(p4?,p5?,p7?)*vx(p5?,p6?,p8?) =
       map(Q1,Q2,p1,p2,p3,p4,p5,p6,p7,p8);

Print +s;
.end
\end{verbatim}
which yields
\begin{verbatim}
   F =
      + map(Q1,Q2,p1,p2,p3,p4,p5,p6,p7,p8)
      + map(Q1,Q2,p1,p7,p4,p3,p8,p6,p2,p5)
      + map(Q1,Q2,p6,p5,p4,p3,p2,p1,p8,p7)
      + map(Q1,Q2,p6,p8,p3,p4,p7,p1,p5,p2)
      + map(Q2,Q1,p3,p2,p1,p6,p5,p4,p8,p7)
      + map(Q2,Q1,p3,p8,p6,p1,p7,p4,p2,p5)
      + map(Q2,Q1,p4,p5,p6,p1,p2,p3,p7,p8)
      + map(Q2,Q1,p4,p7,p1,p6,p8,p3,p5,p2)
     ;
\end{verbatim}

The \texttt{map} function provides a mapping from the old edge labeling
to a new one, which leaves the graph invariant.

\subsection{Output optimization}

Output optimization of polynomials was introduced in \FORM{} 4.1~\cite{Kuipers:2013pba}.
It relies on the idea that the number of arithmetic operations required to evaluate a polynomial can be reduced
by pulling variables outside of brackets (Horner's rule) and by removing common subexpressions.
Finding the best order in which to extract variables (a Horner scheme) is difficult and \FORM{} has different
algorithms for it, namely \cmd{Format O1} to \cmd{Format O4}.

In \FORM{} 4.2 the performance of the output optimization has been increased, by improving the common subexpression detection.
The mode that gives the best results in \FORM{} 4.1, \cmd{Format O3}, has been improved to be less dependent
on a user-given exploration-exploitation constant, based on the work of ~\cite{Ruijl:2013epa} and \cite{Ruijl:2014hha}.

Furthermore, a new algorithm has been added, based on local search methods~\cite{Ruijl:2014spa}. This option
is called \cmd{Format O4} and it generally produces
better results and is faster than the Monte Carlo Tree Search method used in \cmd{Format O3}.
An example is presented below:
\begin{verbatim}
S   a,b,c,d,e,f,g,h,i,j,k,l,m,n;
L   G = (4*a^4+b+c+d + i^4 + g*n^3)^10 +
        (a*h + e + f*i*j + g + h)^8 + (i + j + k + l + m + n)^12;
Format O4,saIter=300;  * use 300 iterations for optimization
.sort
#optimize G
#write "Optimized with Horner scheme: `optimscheme_'"
#write "Number of operations in output: `optimvalue_'"
#clearoptimize
.end
\end{verbatim}

When running with \cmd{tform}, every thread runs its own search and the best value is selected.
This gives better results than using fewer cores with a higher number of optimization iterations, \cmd{saIter}.

\subsection{Automatic series expansion of rational functions}

The \cmd{Polyratfun}, which was introduced in \FORM{} 4.0 to treat multivariate
rational functions as coefficients of terms, now supports automatic expansion
in one variable.
In the example below the rational function represented by the function \cmd{rat}
is expanded with respect to the variable \cmd{ep} up to \cmd{ep\^{}5}:
\begin{verbatim}
S ep;
CF rat;
Polyratfun rat;
L F = rat(ep + 1,ep^2 + 3*ep + 1);
.sort
Polyratfun rat(expand,ep,5);

Print;
.end
\end{verbatim}
The code yields:
\begin{verbatim}
   F =
      rat(1 - 2*ep + 5*ep^2 - 13*ep^3 + 34*ep^4 - 89*ep^5);
\end{verbatim}

All manipulations performed in the expansion mode will truncate
the series quickly. As a result, the expansion mode may be
faster than the unexpanded mode if the coefficients become complicated.
Additionally, it can avoid the intrinsic \cmd{MaxTermSize} restriction of \FORM{}.

\subsection{Textual manipulation on the output by dictionaries}

Occasionally output from computer algebra systems provides input for another
program, which may require suitable translation.
In \FORM{}, the \cmd{Format} statement can be used for preparing the output in
a format of another program, for example, \cmd{Fortran}, \cmd{C} and so on.
\FORM{} version 4.2 gives another way of controlling the output to a certain
extent: \textit{dictionaries}.
A dictionary is a collection of pairs of a \textit{word} and its translation.
The word can be a variable, a number, a function with specific arguments or
a special character like the multiplication sign (\cmd{*}).
The translation can be any string enclosed in double quotes.
For example,
\begin{verbatim}
S x1,x2,x3;
L F = (x1+x2+x3)^2;
.sort
#opendictionary texdict
  #add x1: "x_1"
  #add x2: "x_2"
  #add x3: "x_3"
  #add *: " "
#closedictionary
#usedictionary texdict
#write "%E",F
.end
\end{verbatim}
prints the expression \cmd{F} for the \LaTeX{} math mode:
\begin{verbatim}
   x_3^2 + 2 x_2 x_3 + x_2^2 + 2 x_1 x_3 + 2 x_1 x_2 + x_1^2
\end{verbatim}
Here \cmd{#opendictionary texdict} $\dots$ \cmd{#closedictionary} defines
a dictionary with the name \cmd{texdict}.
Each \cmd{#add} instruction defines a pair of a word and its translation.
The \cmd{#usedictionary} instruction sets the predefined dictionary for
the textual manipulation on the output.
Then, each word in the dictionary is replaced with its translation in
the \cmd{#write} instruction.

\subsection{Spectators}

Sometimes a large number of terms is passed through one or more modules for which
the user knows that they will remain unmodified. Processing these
irrelevant terms will cause overhead, including a potentially heavy sort cost.

The Spectator system can be used to remove terms from the current expression
and to store them in a buffer, called a Spectator file. These terms
will be ignored until they are copied back into an active expression.
In essence, the Spectator system works as a filter.
Below is an example:
\begin{verbatim}
S x;
CF f,g;

CreateSpectator FSpec "Fspec.spec";
L F = f(1);
.sort

#do i=1,10000
  id f(x?) = theta_(10000-x)*f(x + 1) + g(x);  * term blow-up
  if (count(g,1)) ToSpectator FSpec;  * filter the g terms
  .sort:round`i';
#enddo

CopySpectator F1 = FSpec;
.sort
RemoveSpectator FSpec;
.end
\end{verbatim}

The above code takes about 0.25 seconds to run.
If the line with \cmd{ToSpectator} is commented out, it takes about 13 seconds to
run.

\subsection{Zero-dimensional sparse tables as pure functions}

High-level programming languages are mainly classified into two groups:
\emph{imperative} languages and \emph{declarative} languages.
Modern programming languages, however, tend to have aspects of both the groups
and evolve towards multi-paradigm languages, which makes some programming tasks
easier or more intuitive.
\FORM{}, which is clearly imperative, is not an exception.
The users are now allowed to declare tables without any table indices.
Such zero-dimensional (sparse) tables can have function arguments with
wildcarding, which effectively leads to user-defined \emph{pure functions}
and opens a way of functional programming up to some extent.
For example,
\begin{verbatim}
S n;
Table fac(n?pos0_);
Fill fac = delta_(n) + theta_(n) * n * fac(n-1);

L F = fac(5);
Print;
.end
\end{verbatim}
implements a function \cmd{fac} as
\begin{equation}
  \cmd{fac}(n) =
  \begin{cases}
    1 & n = 0, \\
    n \times \cmd{fac}(n-1) & n \ge 1 .
  \end{cases}
\end{equation}
for a non-negative integer $n$, and yields
\begin{verbatim}
   F =
      120;
\end{verbatim}
Of course in this case one can use the more efficient built-in \cmd{fac_} function.%
\footnote{%
  In addition to the performance issue, the user-defined function
  suffers from a restriction with the current implementation of \FORM{}:
  the recursion depth is limited by the stack size.
}

A more complicated example is given below:
\begin{verbatim}
S n,n1,n2;
Table fib(n?int_);
Table fibimpl(n?int_,n1?,n2?);
Fill fib =   theta_(-1-n) * sign_(n+1) * fib(-n)
           + theta_(n-1) * fibimpl(n-2,1,1);
Fill fibimpl =   theta_(-n) * n2
               + thetap_(n) * fibimpl(n-1,n2,n1+n2);

L F = fib(100);
Print;
.end
\end{verbatim}
This computes the 100th Fibonacci number:
\begin{verbatim}
   F =
      354224848179261915075;
\end{verbatim}

\subsection{Expression database using \texorpdfstring{\cmd{ArgToExtraSymbol}}{}}

\FORM{} has tables that map integers (keys) to expressions (values).
An arbitrary expression cannot be a key of a table in a direct way.
However, the new command \cmd{ArgToExtraSymbol} with the \cmd{tonumber} option
gives a way to map an arbitrary expression into an integer, leading to
the use of an expression as a key to a table.

Below we give an example where we effectively build a table storing
\begin{align*}
  \cmd{g(1)*g(2)} &\Longrightarrow \cmd{100} , \\
  \cmd{g(1)*g(2)*g(3)} &\Longrightarrow \cmd{200} ,
\end{align*}
by using the following code:
\begin{verbatim}
S n;
CF f,g;
L F = f(g(1)*g(2))*100 + f(g(1)*g(2)*g(3))*200;

argtoextrasymbol tonumber,f;

B f;
Print;
.sort
CTable sparse,values(1);
FillExpression values = F(f);  * store values in table
Drop F;
.sort

L G = f(g(1)*g(2)) + f(g(1)*g(2)*g(3)) + f(g(1)*g(2)*g(3)*g(4));

argtoextrasymbol tonumber,f;
id f(n?) = f(extrasymbol_(n))*values(n);  * read values in table

Print +s;
.end
\end{verbatim}

The first \cmd{Print} statement gives:
\begin{verbatim}
   F =
       + f(1) * ( 100 )
       + f(2) * ( 200 );
\end{verbatim}

and the final output is
\begin{verbatim}
   G =
       + 100*f(g(1)*g(2))
       + 200*f(g(1)*g(2)*g(3))
       + f(g(1)*g(2)*g(3)*g(4))*values(3)
      ;
\end{verbatim}

\subsection{Partition function}

We have added several new convenient statements and functions to \FORM{} 4.2,
such as \cmd{Transform dedup/addargs/mulargs}, \cmd{id_}, \cmd{perm_} and
the \cmd{occurs} condition in the \cmd{If} statement.
Here we showcase the partitions function, which generates all partitions of a list of arguments into
$n$ parts. Each part consists of a function name and a size.
This function exploits symmetries of the arguments to make sure that no partition
is generated twice. Instead, a combinatorial prefactor is computed.

For example, to partition eight elements into three partitions, where
the first part is a function \cmd{f1} with three arguments,
the second part a function \cmd{f1} with two arguments,
and the third part a function \cmd{f2} with three arguments, we write:
\begin{verbatim}
S x1,x2,x3;
CF f1,f2;
L F = partitions_(3,f1,3,f1,2,f2,3,x1,x1,x1,x1,x2,x2,x2,x3);

Print +s;
.end
\end{verbatim}

\begin{verbatim}
   F =
      + 18*f1(x1,x1)*f1(x1,x1,x2)*f2(x2,x2,x3)
      + 6*f1(x1,x1)*f1(x1,x1,x3)*f2(x2,x2,x2)
      + 36*f1(x1,x1)*f1(x1,x2,x2)*f2(x1,x2,x3)
      + 36*f1(x1,x1)*f1(x1,x2,x3)*f2(x1,x2,x2)
      + 6*f1(x1,x1)*f1(x2,x2,x2)*f2(x1,x1,x3)
      + 18*f1(x1,x1)*f1(x2,x2,x3)*f2(x1,x1,x2)
      + 12*f1(x1,x1,x1)*f1(x1,x2)*f2(x2,x2,x3)
      + 4*f1(x1,x1,x1)*f1(x1,x3)*f2(x2,x2,x2)
      + 12*f1(x1,x1,x1)*f1(x2,x2)*f2(x1,x2,x3)
      + 12*f1(x1,x1,x1)*f1(x2,x3)*f2(x1,x2,x2)
      + 72*f1(x1,x1,x2)*f1(x1,x2)*f2(x1,x2,x3)
      + 36*f1(x1,x1,x2)*f1(x1,x3)*f2(x1,x2,x2)
      + 18*f1(x1,x1,x2)*f1(x2,x2)*f2(x1,x1,x3)
      + 36*f1(x1,x1,x2)*f1(x2,x3)*f2(x1,x1,x2)
      + 36*f1(x1,x1,x3)*f1(x1,x2)*f2(x1,x2,x2)
      + 18*f1(x1,x1,x3)*f1(x2,x2)*f2(x1,x1,x2)
      + 36*f1(x1,x2)*f1(x1,x2,x2)*f2(x1,x1,x3)
      + 72*f1(x1,x2)*f1(x1,x2,x3)*f2(x1,x1,x2)
      + 12*f1(x1,x2)*f1(x2,x2,x3)*f2(x1,x1,x1)
      + 36*f1(x1,x2,x2)*f1(x1,x3)*f2(x1,x1,x2)
      + 12*f1(x1,x2,x2)*f1(x2,x3)*f2(x1,x1,x1)
      + 12*f1(x1,x2,x3)*f1(x2,x2)*f2(x1,x1,x1)
      + 4*f1(x1,x3)*f1(x2,x2,x2)*f2(x1,x1,x1)
     ;
\end{verbatim}

All options for \cmd{partitions_} can be found in the manual~\cite{Manual}.

\section{Summary}
\label{sec:summary}

We have shown some new features of \FORM{} 4.2. With these features we hope 
to make the usage of \FORM{} easier and to have improved performance. This 
version of \FORM{} is the version used by the \textsc{Forcer}~\cite{FORCER} 
package and its supporting software.

\section*{Acknowledgements}

We would like to thank everyone who has given feedback and reported bugs, in particular Joshua Davies.
This work is supported by the ERC Advanced Grant no. 320651, ``HEPGAME''.

\bibliographystyle{JHEP}
\bibliography{myref}

\providecommand{\url}[1]{#1}\providecommand{\href}[2]{#2}\begingroup\raggedright\begin{thebibliography}{10}

\bibitem{Vermaseren:2000nd}
J.A.M.~Vermaseren, \emph{{New features of FORM}},
  \href{https://arxiv.org/abs/math-ph/0010025}{{\ttfamily math-ph/0010025}}.

\bibitem{Kuipers:2012rf}
J.~Kuipers, T.~Ueda, J.A.M.~Vermaseren and J.~Vollinga, \emph{{FORM version
  4.0}}, \href{https://doi.org/10.1016/j.cpc.2012.12.028}{\emph{Comput. Phys.
  Commun.} {\bfseries 184} (2013) 1453}
  [\href{https://arxiv.org/abs/1203.6543}{{\ttfamily arXiv:1203.6543}}].

\bibitem{Changelog}
\emph{{FORM 4.2 Changelog}},
  \url{https://github.com/vermaseren/form/wiki/Release-Notes---Form-4.2.0#changelog---form-420-2017-07-06}.

\bibitem{FORCER}
B.~Ruijl, T.~Ueda and J.A.M.~Vermaseren, \emph{{Forcer, a FORM program for the
  parametric reduction of four-loop massless propagator diagrams}},
  \href{https://arxiv.org/abs/1704.06650}{{\ttfamily arXiv:1704.06650}}.

\bibitem{Repository}
\emph{{FORM Source Repository}},  \url{https://github.com/vermaseren/form}.

\bibitem{Tentyukov:2007mu}
M.~Tentyukov and J.A.M.~Vermaseren, \emph{{The Multithreaded version of FORM}},
  \href{https://doi.org/10.1016/j.cpc.2010.04.009}{\emph{Comput. Phys. Commun.}
  {\bfseries 181} (2010) 1419}
  [\href{https://arxiv.org/abs/hep-ph/0702279}{{\ttfamily hep-ph/0702279}}].

\bibitem{Issues}
\emph{{FORM Issue Tracker}},  \url{https://github.com/vermaseren/form/issues}.

\bibitem{Manual}
\emph{{FORM 4.2 Reference Manual}},
  \url{https://github.com/vermaseren/form/releases/download/v4.2.0/form-4.2.0-manual.pdf}.

\bibitem{jvLL2016}
F.~Herzog, B.~Ruijl, T.~Ueda, J.A.M.~Vermaseren and A.~Vogt, \emph{{FORM,
  Diagrams and Topologies}}, {\emph{PoS} {\bfseries LL2016} (2016) 073}
  [\href{https://arxiv.org/abs/1608.01834}{{\ttfamily arXiv:1608.01834}}].

\bibitem{Kuipers:2013pba}
J.~Kuipers, T.~Ueda and J.A.M.~Vermaseren, \emph{{Code Optimization in FORM}},
  \href{https://doi.org/10.1016/j.cpc.2014.08.008}{\emph{Comput. Phys. Commun.}
  {\bfseries 189} (2015) 1} [\href{https://arxiv.org/abs/1310.7007}{{\ttfamily
  arXiv:1310.7007}}].

\bibitem{Ruijl:2013epa}
B.~Ruijl, J.~Vermaseren, A.~Plaat and H.J.~van~den~Herik, \emph{{Combining
  Simulated Annealing and Monte Carlo Tree Search for Expression
  Simplification}}, {\emph{Proceedings of ICAART Conference 2014} {\bfseries 1}
  (2014) 724} [\href{https://arxiv.org/abs/1312.0841}{{\ttfamily
  arXiv:1312.0841}}].

\bibitem{Ruijl:2014hha}
B.~Ruijl, J.~Vermaseren, A.~Plaat and J.v.d.~Herik, \emph{{HEPGAME and the
  Simplification of Expressions}},
  \href{https://arxiv.org/abs/1405.6369}{{\ttfamily arXiv:1405.6369}}.

\bibitem{Ruijl:2014spa}
B.~Ruijl, A.~Plaat, J.~Vermaseren and J.~van~den~Herik, \emph{{Why Local Search
  Excels in Expression Simplification}},
  \href{https://arxiv.org/abs/1409.5223}{{\ttfamily arXiv:1409.5223}}.

\end{thebibliography}\endgroup

\end{document}